\documentclass[11pt]{article}
\usepackage{epsfig}
\def\d{\partial}
\def\db{\bar\partial}
\def\t{\theta}
\def\tb{\bar\theta}
\usepackage{oldgerm}
\input{amssym}
\def\be{\begin{equation}}
\def\ee{\end{equation}}

\begin{document}
\title{Spatial Asymmetric Two dimensional Continuous Abelian Sandpile Model}
\author{N. Azimi-Tafreshi\footnote{email:azimi@physics.sharif.ir}, H. Dashti-Naserabadi, S. Moghimi-Araghi\footnote{email: samanimi@sharif.edu}\\
Department of Physics, Sharif University of Technology,\\ Tehran,
P.O.Box: 11155-9161, Iran} \date{} \maketitle

\begin{abstract}
We insert some asymmetries in the continuous Abelian sandpile
models, such as directedness and ellipticity. We analyze probability
distribution of different heights and also find the field theory
corresponding to the models. Also we find the fields associated with
some height variables.

  \vspace{5mm}%
\end{abstract}

\section{Introduction}

After the seminal work of Bak, Tang and Wiesenfeld \cite{BTW},
sandpile models have been found to be a very helpful framework to
study Self-Organized Criticality (SOC).  SOC is believed to be the
underlying reason for the scaling laws seen in a number of natural
phenomena \cite{Jensen}. Further investigations showed that BTW
sandpile model has an Abelian property and was renamed Abelian
Sandpile Model (ASM) \cite{Dhar}. The model is still the simplest,
most studied model of SOC, in which many analytical results has been
derived. For a good review see ref. \cite{DharRev}.

Extensive work has been done on this model, analytical and
computational. To name a few, one can list the works by Dhar
\cite{Dhar,MajDharJPhysA}, where the probabilities of some height
cluster were computed. In \cite{Dhar}, Dhar computed the number of
recurrent configurations and showed that all occur with equal
probability. Later, in \cite{MajDharJPhysA}, Majumdar and Dhar
calculated the probabilities of occurrence of some specific
clusters, known as Weakly Allowed Clusters (WAC's). The simplest of
these clusters is one-site height one cluster. The probabilities of
other one-site clusters with height above 1 was computed in ref.
\cite{Prizh}. There are many other analytical results, among them
one can mention the results on boundary correlations of height
variables and effect of boundary conditions
\cite{Ivashkevich94,Jeng05-1,Jeng05-2,Jeng04,Ruelle02,Ruelle07}, on
presence of dissipation in the model \cite{Jeng05-2,Jeng04,PirRu04},
on field theoretical approaches \cite{Jeng05-2,MRR,JPR,MaRu,MN}, on
finite size corrections \cite{MajDharJPhysA,Ruelle02} and many other
results\cite{DharRev}.

In ASM, the height variables take only integer values. However there
are some other models in which the height variables can take real
values \cite{Zhang,Gabrielov,GLJ,Tsuchia,ALM}. In Zhang model, which
is not Abelian, when a site topples all of its energy (height) is
distributed among its neighbors. On the other hand the model
introduced by Gabrielov, called Abelian Avalanche Model (AA Model)
and the model introduced by Ghaffari et. al. have the Abelian
property. In \cite{ALM} a continuous version of  ASM is considered
on a square lattice and many of its different properties are
studied. When we let the heights be real number, naturally the
components of toppling matrix could be real too. In this paper we
consider continuous ASM \cite{ALM} and change different components
of toppling matrix. We will do it in a way to add spatial
asymmetries to the model. Two specific asymmetries are considered,
one is to introduce a preferred direction. The second is to
introduce elliptical asymmetry. The first one is related to directed
sandpile model and the second, as we shall see results naturally
from the original BTW model. We have derived the probability
distribution of height variables and also avalanche distributions.
Also the field theory associated with these models are considered.
The new models are still critical and show scaling relations.

 In the next section we first
review the Continuous ASM. Next we consider the effects of a
perturbation that introduces a preferred direction in the model. In
the forth section we introduce ellipticity in the model and derive
many properties of such a system.

\section{Continuous Abelian Sandpile Model }

The Continuous Abelian Sandpile Model (CASM) on a square lattice is
defined in \cite{ALM}. Consider a $L\times L$ square lattice. To
each site $u=(i,j)$ a {\it continuous} height variable, $h(u)$ is
assigned, where without loss of generality we assume these variables
are in [0,4) (in the original paper it was taken to be in [0,1) ).
The evolution rules consist of the following rules:

1) At each time step, a site is selected randomly and an amount of
sand is added to it and other height variables are unchanged. The
amount of sand added to the site is a random real number in the
interval $[p,q]\in[0,4)$. The distribution function could be any
function, but for simplicity we take it to be uniform on the
interval $[p,q]$. If height remains below four, the new
configuration is stable and we go to the next time step.

2) If height of the sand at that site becomes equal to or greater
than four, the site becomes unstable and topples in the following
way: it gives an amount of sand with height one to either of its
neighbors. In other words $h(v)\rightarrow h(u)-\Delta^C(u,v)$ for
all $v$, where $\Delta^C(u,v)$ is the toppling matrix of the CASM
and is defined as
\begin{eqnarray}
\Delta^C (u,v) \left\{
\begin{array}{cc}
4 & u=v\\
-1 &  |u-v|=1\\
0 & {\rm otherwise}
\end{array}
\right.
\end{eqnarray}

So if the height of a site is $4+\alpha$, after toppling its height
will be $\alpha$ and the heights of neighbors will be added by the
amount 1. As a result of this toppling, some of the neighbors may
become unstable and an avalanche may occur.

It has been shown in \cite{ALM} that there is a (many to one)
mapping from configurations in CASM to configurations in ASM, which
preserves the dynamics, therefore the new model reproduce most of
the interesting features of ASM. Also it is shown that the
probability distribution is piecewise constant. The probability of
finding a site with height $h\in[k-1,k)$ with $k=1,2,3,4$ is equal
with $p(k)$, the probability of finding a site with height $k$ in
the usual ASM. The claims were proved and were in agreement with
simulations.

As the heights are continuous, we can take other integer parameters
of the model to be real. This has been done to the dissipation
parameter in the model\cite{ALM}. The probability distribution is
found when there exist dissipation with different (real) values in
\cite{ALM}. Yet there are some other parameters which are integer in
the original model, but may be taken to be real in the new model. In
this paper we consider the amount of sand transferred to the
neighbors and let it take different real values and see how the
properties of the model are changed.

\section{CASM with preferred direction}
One of the easiest modifications to the toppling matrix is to
introduce a preferred direction, that is; sand grains are more
likely to move to left rather than right. Such modification is also
studied in the usual ASM \cite{DharRamas}, but as the toppling
matrix components have integer values, the only possibility is to
send two sand grains to left and nothing to right. The same can be
done in the vertical direction. If this is the case, then the
toppling matrix will be upper triangular and all the configurations
will be recurrent \cite{DharRamas}. However, in our model we are
able to control the amount of directedness: when a site is toppled,
then $1+\epsilon$ grains of sand move to left and $1-\epsilon$
grains of sand move to right. Clearly  $\epsilon=0$ corresponds to
the usual ASM and $\epsilon=1$ corresponds to the case mentioned
above, provided we add both horizontal and vertical directedness.

\begin{figure}[t]
\begin{picture}(200,230)(0,0)
\includegraphics{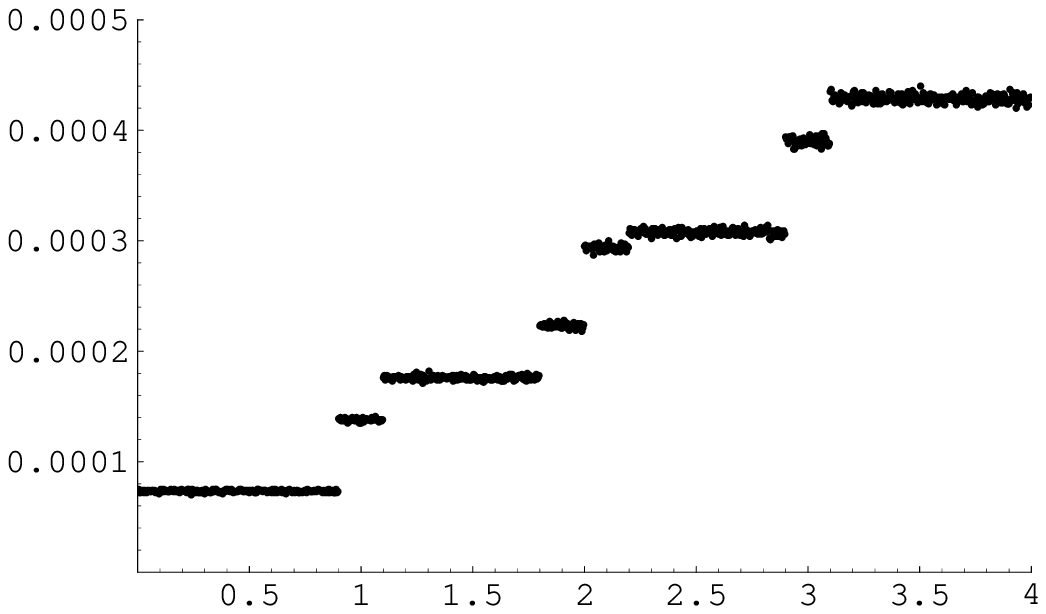}
\includegraphics{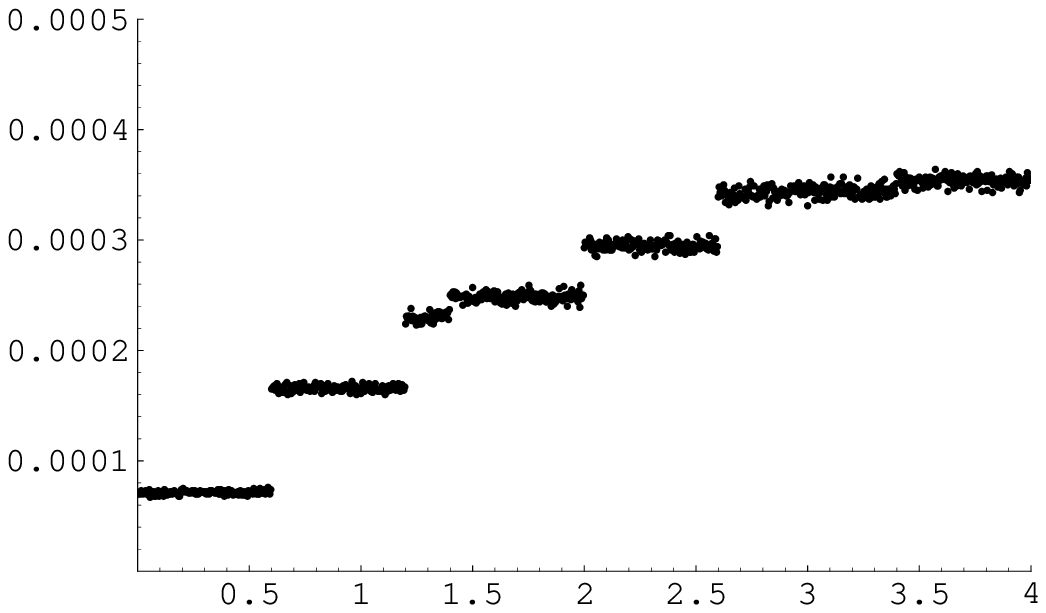} \includegraphics{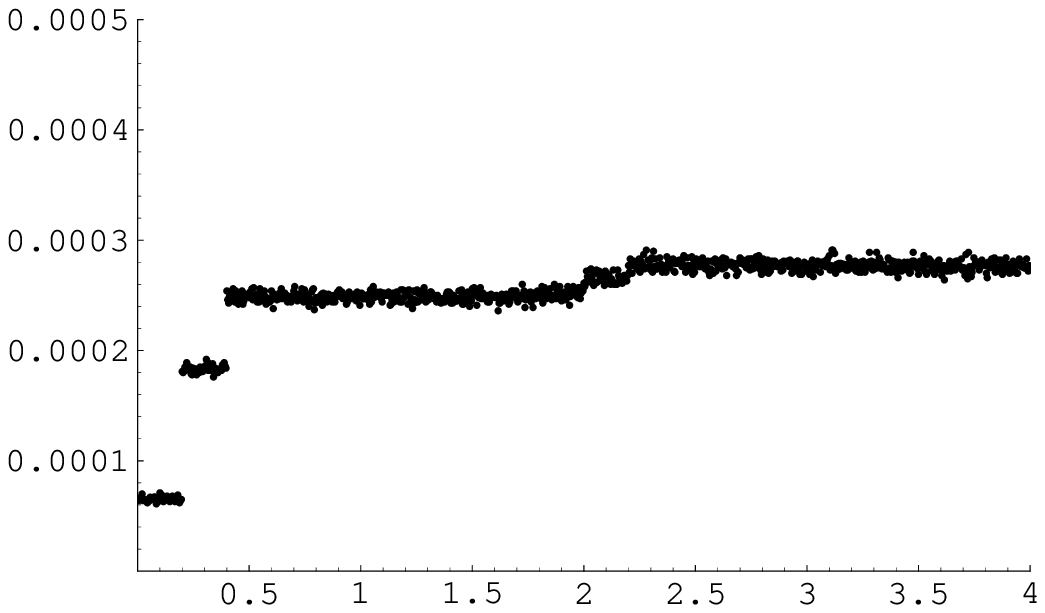} \includegraphics{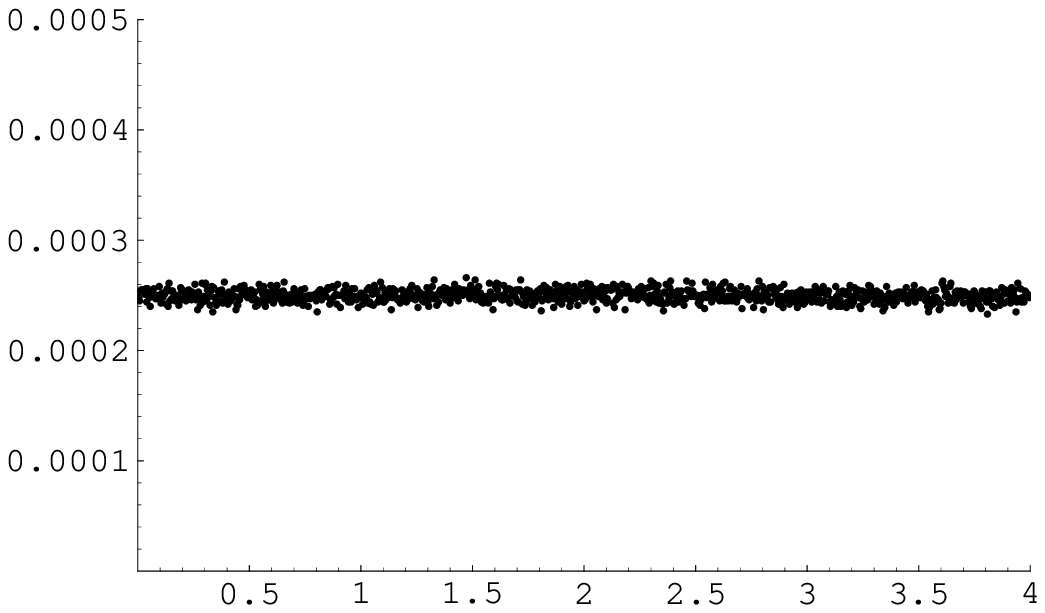}
\put(104,120){$\epsilon=0.1$}\put(304,120){$\epsilon=0.4$}
\put(104,0){$\epsilon=0.8$}\put(304,0){$\epsilon=1.0$}
\end{picture}
\caption{The probability density profile of height variables}
\end{figure}

Several properties of the model can be studied, the probability
distribution, the corresponding action, the avalanche distributions
etc. The probability distribution of finding different heights is
sketched in Figure 1 for some different values of $\epsilon$ when
both directions are directed. As we see, still we have the same
step-like pattern as in the case of undirected CASM. Some new small
steps have appeared which have the length proportional to
$\epsilon$. As $\epsilon$ becomes larger, these steps become longer
and little by little the original pattern fades away. At
$\epsilon=1$ probability distribution becomes uniform, as it is
expected \cite{DharRamas}.

Also the avalanche distribution is studied in this model. System
sizes from 64 to 1024 has been investigated. Starting with a lattice
of randomly distributed heights $h\in[1,4)$, the system is evolved
to reach steady state. After that, we begin the measurements. The
measurements are averaged over about $10^7$ avalanches. We studied
two characteristics of an avalanche (though we bring the result of
one of them): the total number of toppling events is called the size
$s$ of an avalanche and the number of distinct toppled lattice sites
which is denoted by $s_d$, . Because a particular lattice site may
topple several times, the number of toppling events exceeds the
number of distinct toppled lattice sites, i.e., $s>s_d$, except when
the system is fully directed in both directions in which case we
have $s=s_d$. In the critical steady state the corresponding
probability distributions should obey power-law behavior
characterized by exponents $\tau_s$ and $\tau_d$:
\begin{eqnarray}
P(s)\propto s^{-\tau_s},\\
P(s_d)\propto s_d^{-\tau_d}
\end{eqnarray}

\begin{figure}[b]
\begin{picture}(200,100)(0,0)
\includegraphics{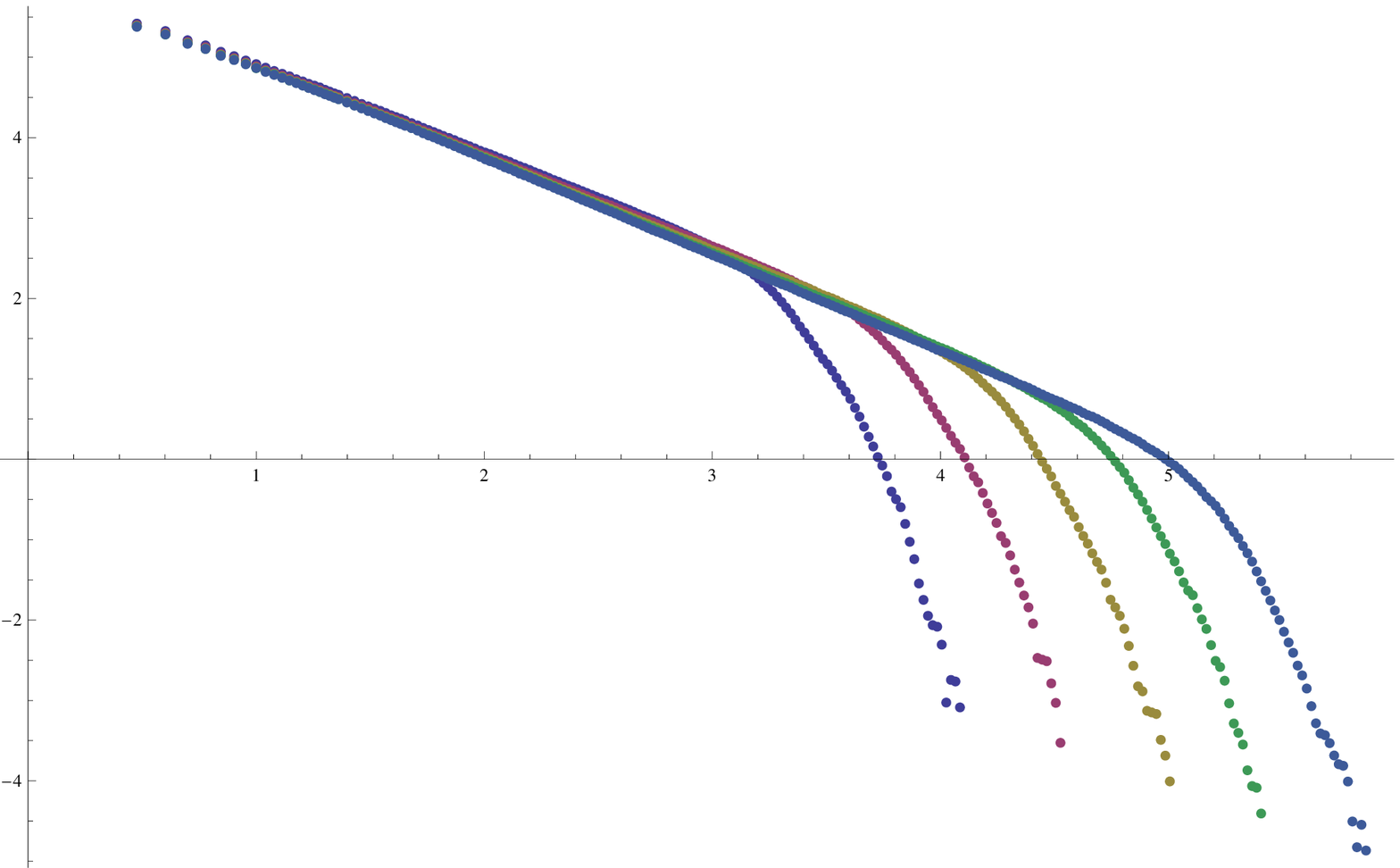}
\includegraphics{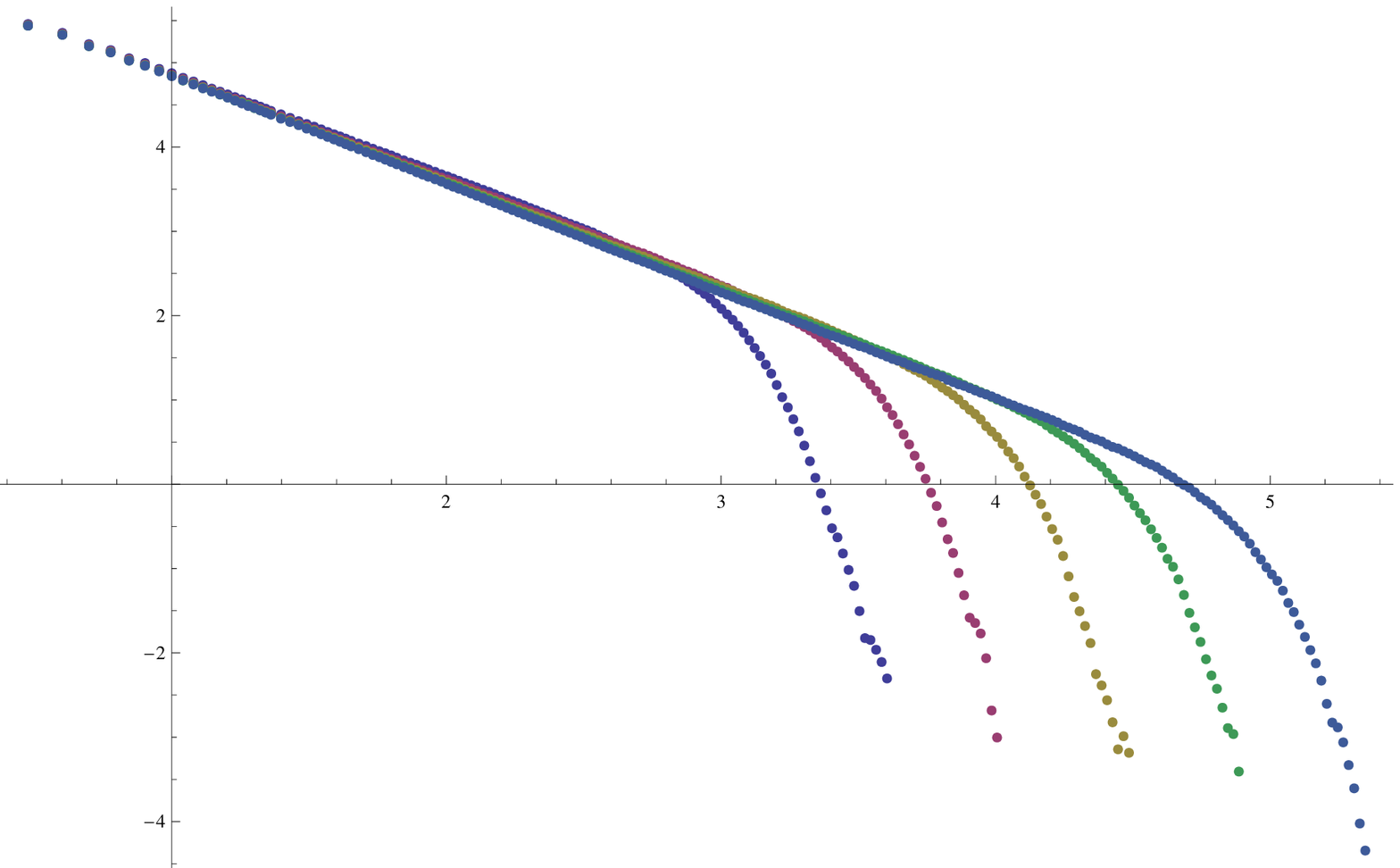}
\includegraphics{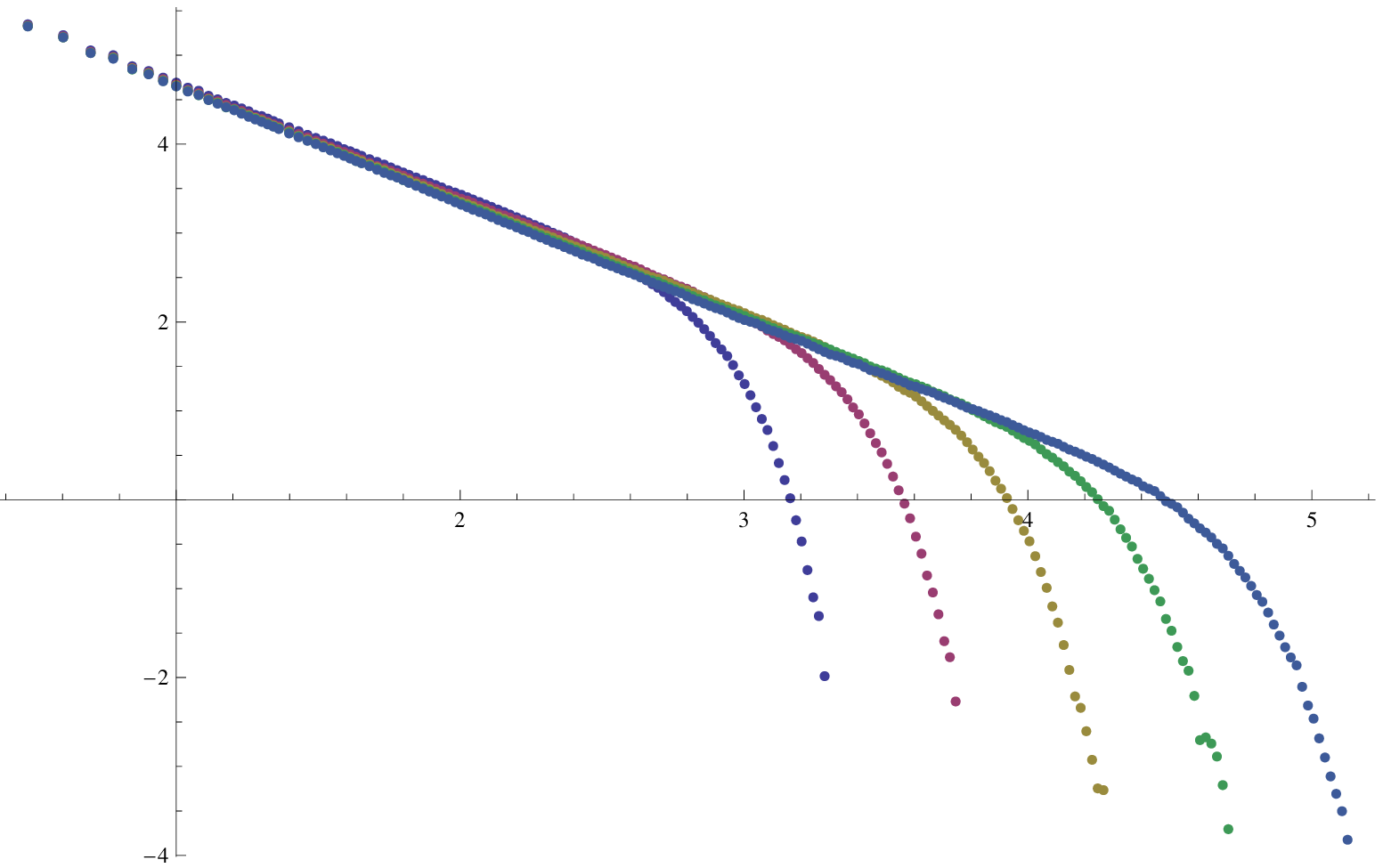}
\put(55,0){$\epsilon=0.1$}\put(220,0){$\epsilon=0.4$}\put(383,0){$\epsilon=1.0$}
\end{picture}
\caption{Avalanche size distribution for $\epsilon=0.1,\,0.4,\,1.0$
and for lattice sizes $L=64,\,128,\,256,\,512,\,1024$.}
\end{figure}
\begin{figure}[t]
\begin{picture}(200,80)(20,0)
\includegraphics{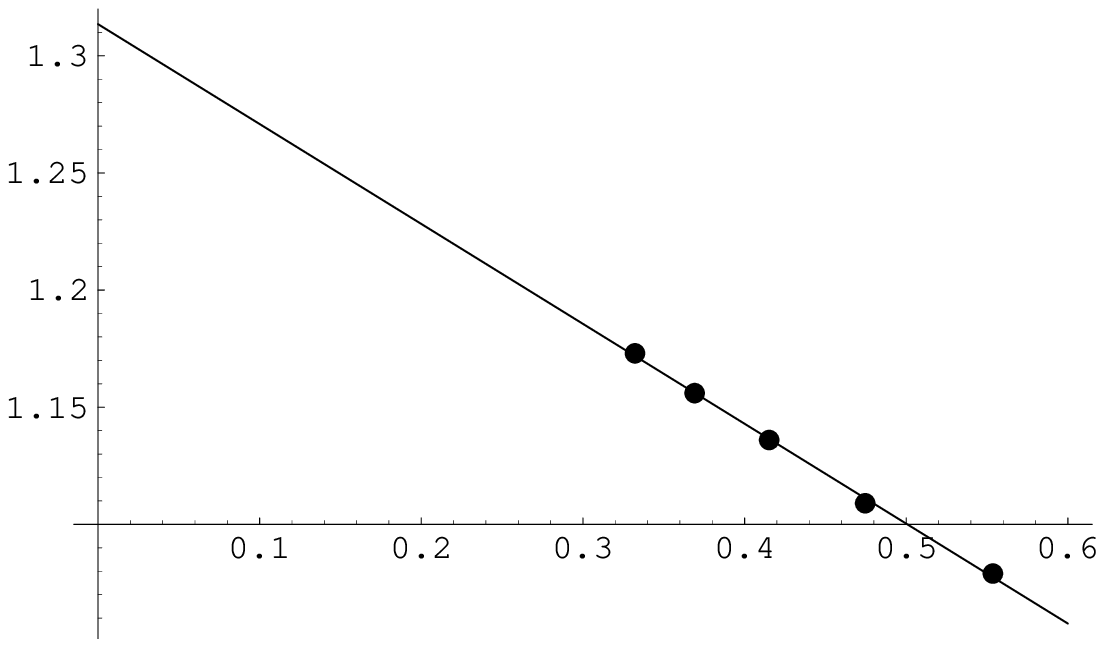}
\includegraphics{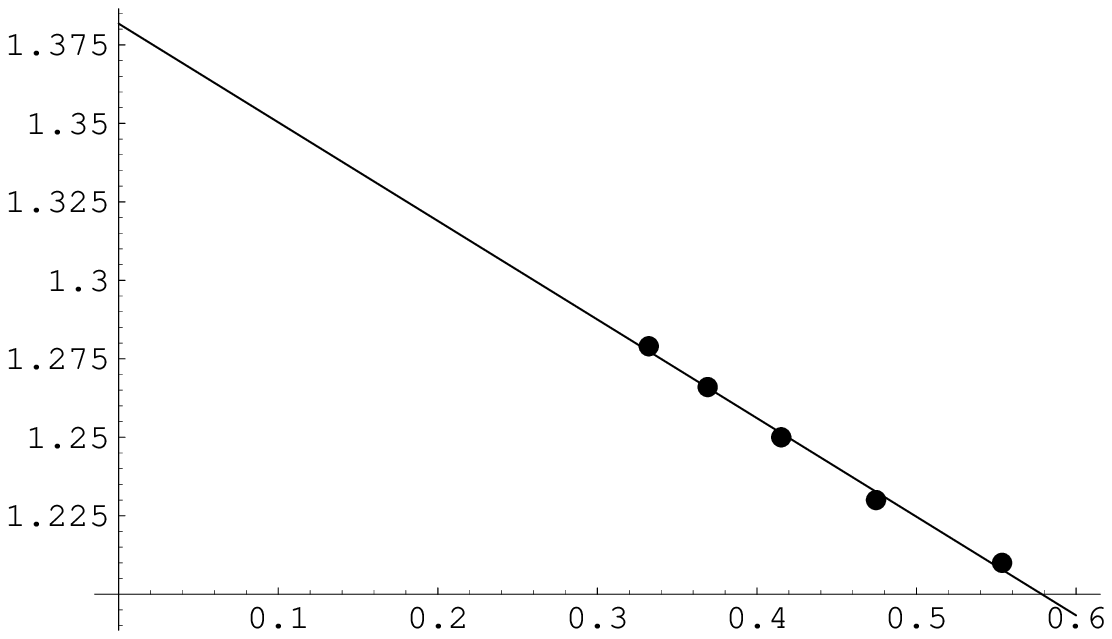}
\put(50,80){$\tau_s(L)$}\put(250,80){$\tau_s(L)$}
\put(205,25){$1/\log L$}\put(415,25){$1/\log L$}
\put(130,0){$\epsilon=0.1$}
\put(330,0){$\epsilon=0.4$}
\end{picture}
\caption{The exponents $\tau_s(L)$ is a linear function of $1/\log
L$. The intersection with vertical axis gives $\tau_s(\infty)$}
\end{figure}
Figure 2 displays the obtained results for the distribution $P(s)$
for different system sizes and three different values of
$\epsilon=0.1,\, 0.4,\,1.0$ when the both directions are directed. A
power-law fit to the straight portion of these curves yields the
exponents $\tau_s(L)$. Figure 3 shows a plot of the exponents
$\tau_s(L)$ vs $1/\log L$ for $\epsilon=0.1$ and $0.4$. This allows
us to extract the exponent $\tau_s(\infty)$. As example, for the
cases  shown in Figure 3, it turns out to be: $\tau_s(\infty)=1.31
\pm 0.08$ for  $\epsilon=0.1$ and $\tau_s(\infty)=1.38\pm 0.08$ for
$\epsilon=0.4$, though the error bars are quite big and we do not
have reliable results at hand. It is worth to mention that for small
values of $s$, the magnitude of slope is less comparing with the one
associated with greater values of $s$ which is due to finite size of
lattice. It is seen that still the system is critical and
self-organized. There is no characteristic length in the model and
finite size scaling is clearly observed from Figure 2. The exponent
$\tau_s$ seems to increase as $\epsilon$ becomes greater which could
be expected from the known behavior of directed sandpile model.

In the continuum limit, we may assign an action to the theory. As in
the steady state, only recurrent configurations appear, and they
appear with equal probability, the partition function is simply
equal with the total number of recurrent configurations, that is,
the determinant of toppling matrix. Such determinants could be
written in terms of integration over grassman variables:
\begin{equation}
\det\Delta=\int d\t_i d\tb_i \exp\left(\t_i \Delta_{ij}\tb_j\right).
\end{equation}
In this way the action of the theory may be obtained. If lattice
spacing is small, one can go to continuum limit and find a field
theoretic action for the theory. In the case of usual ASM the action
turns out to be the $c=-2$ logarithmic conformal theory action
\cite{IvashPriz,MRR}:
\begin{equation}\label{c-2action}
S_{c=-2}\propto\int \db \theta \d \tb
\end{equation}
 For the theory defined above, the action is more or less the same,
 but there are some added terms due to directedness:
\begin{equation}
S_{\rm dir.}\propto\int \left(\db \theta \d \tb +  \t (a \d+\bar{a}
\db)\tb\right),
\end{equation}
where $a$ is a constant proportional to $\epsilon$ with dimension of
length.  This new term grows under renormalization group, therefore
the large scale properties of the model is given by the fully
directed model.

\section{Elliptical asymmetry}

In the previous section, it was supposed that the lattice has a
preferred direction. The field associated with this deformation was
a relevant one and hence we expected the theory show different
characteristics on large scales. Also the toppling matrix was not
symmetric: the amount of sand transferred from site $i$ to the site
$j$ is not necessarily equal with the amount of sand transferred
from $j$ to $i$. Therefore one can not apply the burning test to the
model. However it is possible to modify the toppling matrix in the
following way so that it remains symmetric and, as we shall see, the
added term to the action would not be relevant.

We assume that the vertical links can carry an amount of sand equal
to $1-\epsilon$ and the horizontal links can carry an amount of sand
equal with $1+\epsilon$ ($0\leq\epsilon \leq 1$). In this way the
total amount of sand removed from a toppled site is 4 and the mass
of sand is conserved. As we see the new toppling matrix is
symmetric, but in real space, $x$ direction is quite different from
$y$ direction, in fact; we have added some ellipticity to the system
and therefore it does not have full rotational symmetry, however
this deformation preserves the scale symmetry.

If we go back to the original BTW paper \cite{BTW}, we observe that
elliptical asymmetry naturally could arise in their model. In their
paper, BTW first define a one dimensional model and using a specific
transformation, express it in some other parameters. The same is
done to a two dimensional version of the model to arrive at BTW
model, however it could easily be observed that if we apply the same
transformation to the two dimensional model, we will not arrive at
the BTW model, instead we arrive at a model with elliptical
asymmetry. Let's explain it in more details. In the one dimensional
model, an ordered array of heights $z_i$ is considered. Sand grains
enter from left and leave from right. If the difference of the
heights of two neighbors become more than two a sand grain moves to
right. The slope parameters $h_i$ are defined as $h_i=z_i-z_{i+1}$.
Then as result of "toppling" at site $i$ we will have
$h_i\rightarrow h_i-2$ and $h_{i\pm 1}\rightarrow h_{i\pm 1}+1$. The
generalization to two dimensions would be to define the slope
parameters as $h_i=2z_i-z_{i,R}-z_{i,D}$, where $z_{i,R}$ and
$z_{i,D}$ are the height variables right and down to the site $i$
respectively. The dynamics would be if the slope parameter of a cite
(mean value of the slope there) is more than a threshold, one sand
grain moves to right and one sand grain moves downward. The dynamics
in terms of $h$ variables would be $h_i\rightarrow h_i-4$,
$h_{i,L;R;D;U}\rightarrow h_{i,L;R;D;U}+2$ and
$h_{i,RU;LD}\rightarrow h_{i,RU;LD}-1$, where $h_{i,RU;LD}$ are the
slope parameters of right-up and left-down next nearest neighbors of
the site $i$. As we see there is an asymmetry in the two diagonal
directions and the system does not have full rotational symmetry.

\begin{figure}[t]
\begin{picture}(200,100)(0,0)
\includegraphics{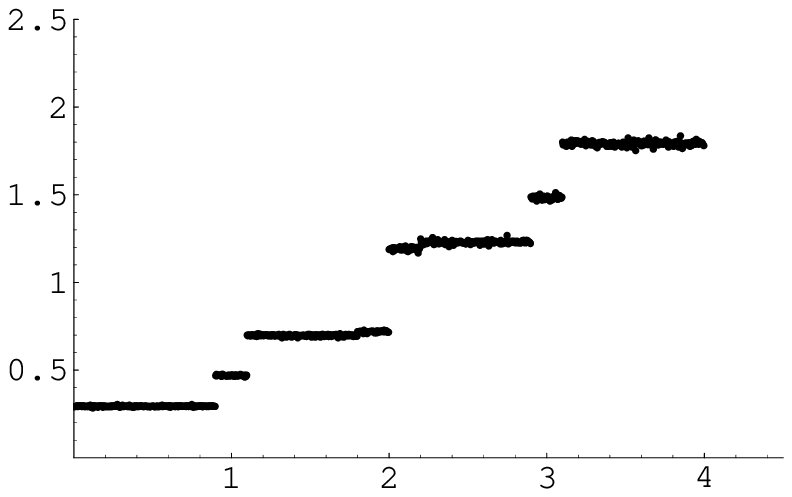} \includegraphics{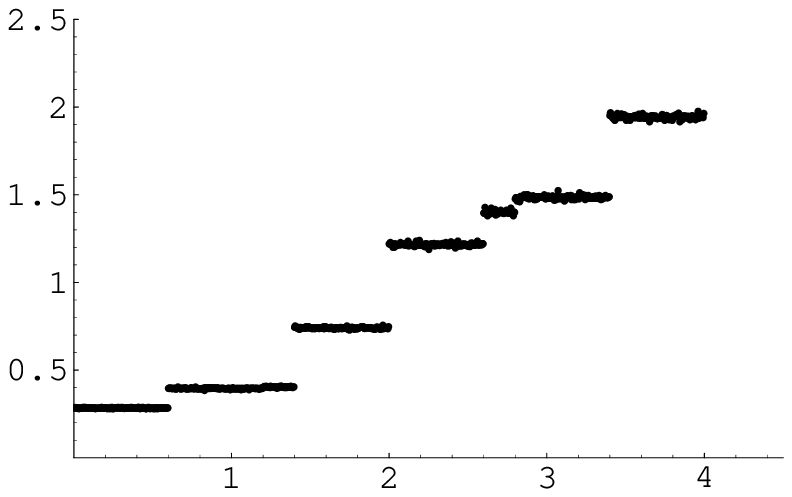} \includegraphics{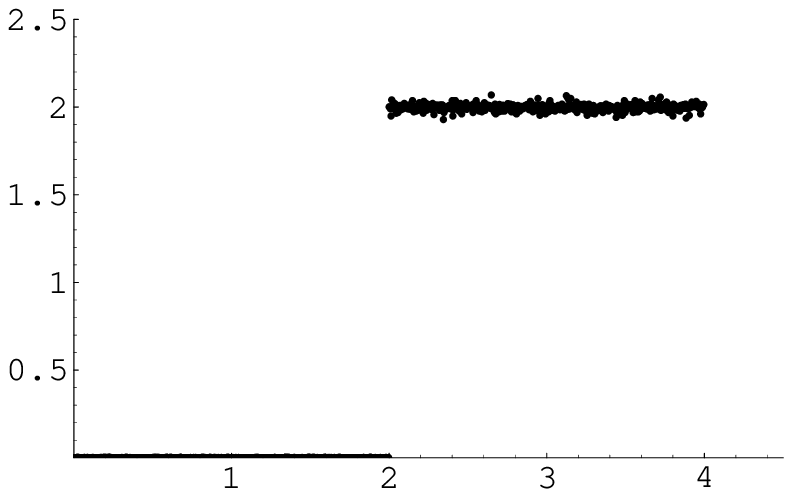}
\put(55,0){$\epsilon=0.1$}\put(220,0){$\epsilon=0.4$}\put(383,0){$\epsilon=1.0$}
\end{picture}
\caption{The probability density profile of height variables in
elliptical asymmetry}
\end{figure}
The probability distribution of finding different heights is
sketched in Figure 4 for $\epsilon\in\{0.1,0.4,1.0\}$. The
simulation is done on a $1024\times 1024$ lattice averaging over
$10^7$ samples. We still have major and minor steps when $\epsilon$
is small. As $\epsilon$ becomes greater, the minor steps become
larger and at $\epsilon=1$ we have a two-step graph, which indicates
that the model is now effectively a one dimensional one.

Also the avalanche distribution is studied in this model. Just as
the case of directed CASM, system sizes from 64 to 1024 has been
investigated. Again, starting with a lattice of randomly distributed
heights $h\in[1,4]$, the system is evolved to reach steady state.
After that, the measurements are begun. The measurements are
averaged over about $10^7$ avalanches. Probability distribution of
avalanche size is sketched in Figure 5 for $\epsilon=0.1,\, 0.4$.
The case of $\epsilon=1.0$ is totally different, as it corresponds
to one dimensional ASM.
\begin{figure}[t]
\begin{picture}(200,100)(0,0)
\includegraphics{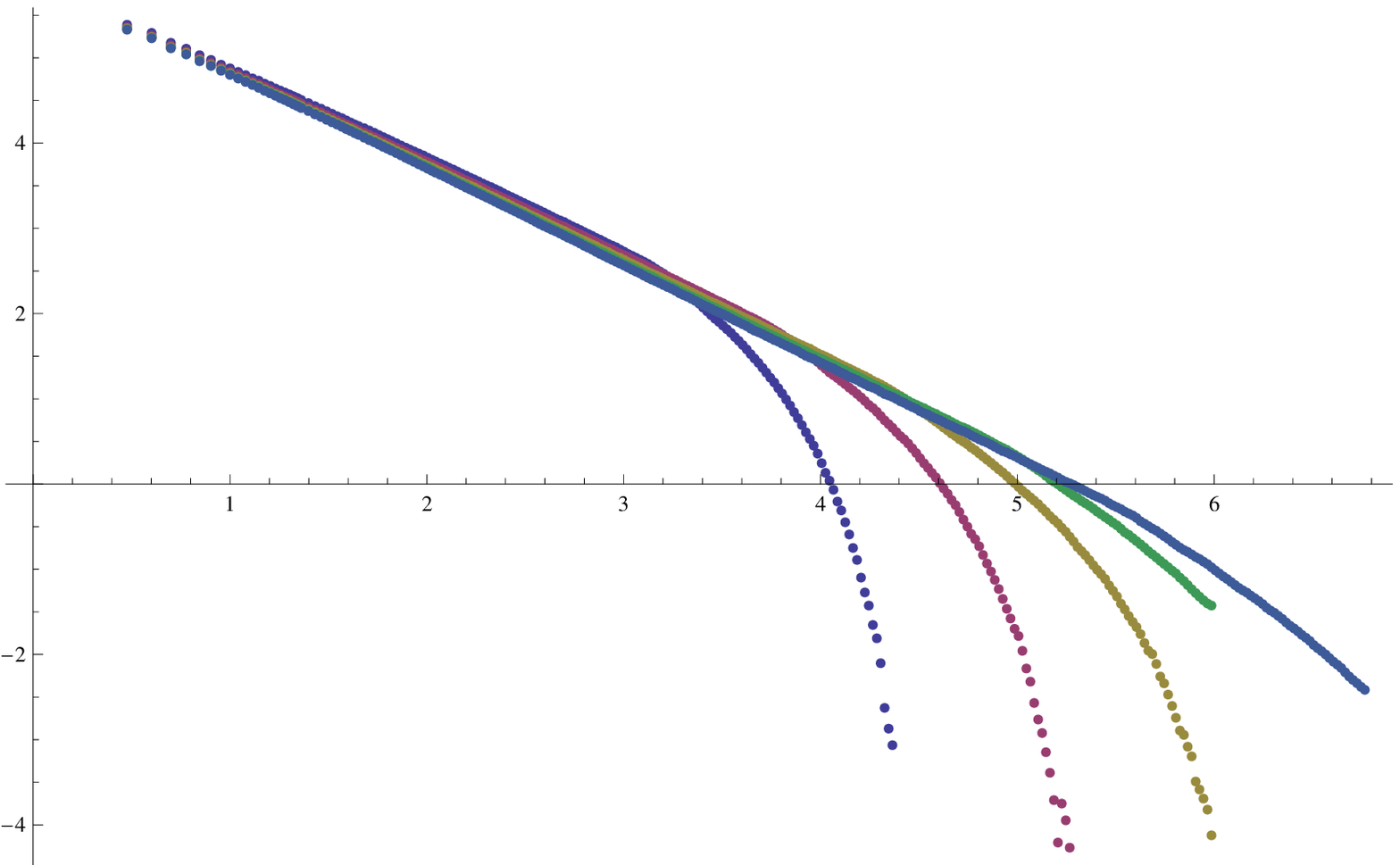} \includegraphics{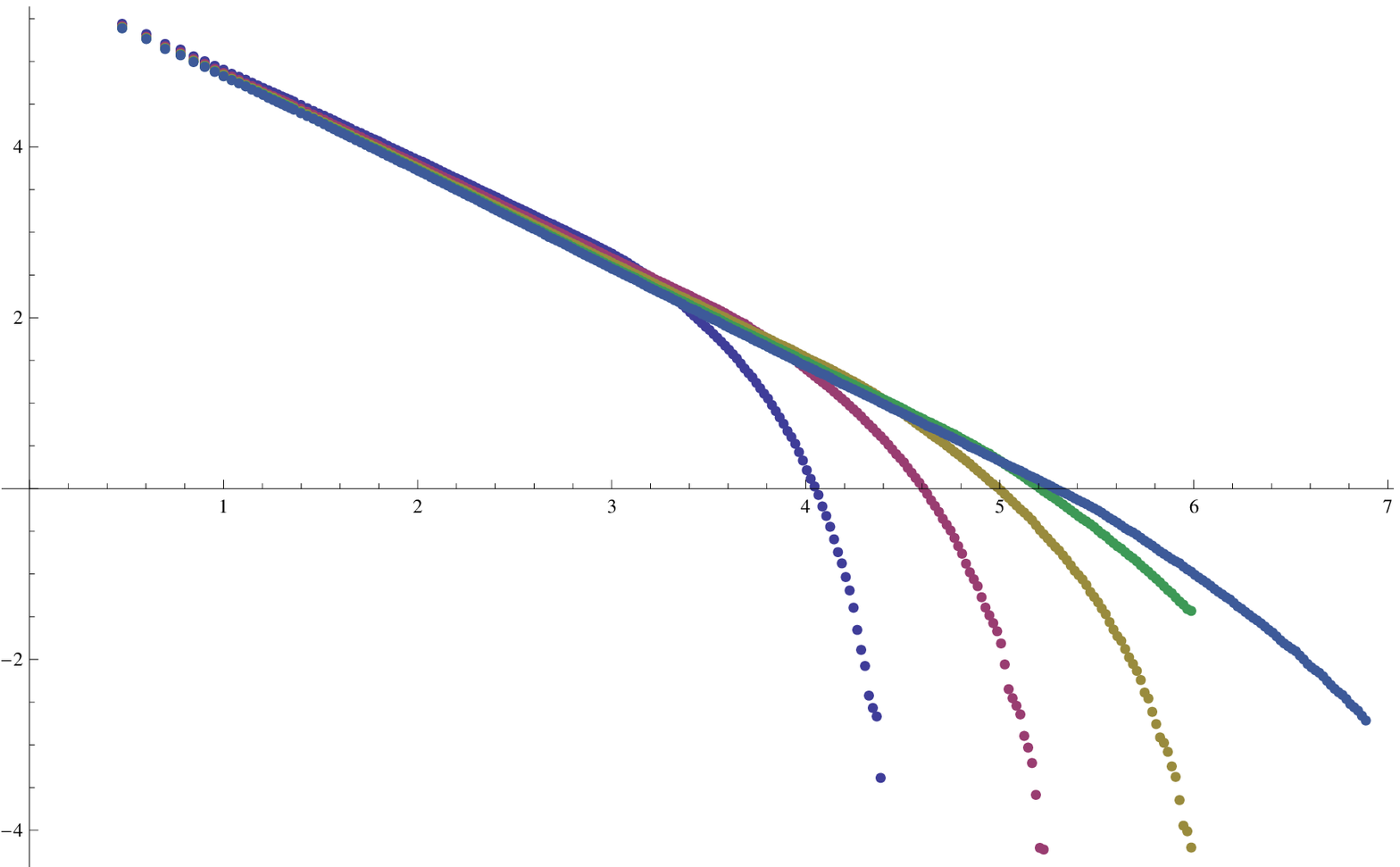}
\put(100,3){$\epsilon=0.1$}\put(320,3){$\epsilon=0.4$}
\end{picture}
\caption{The avalanche size distribution in a system with elliptical
asymmetry.}
\end{figure}

Using The same tools, the slope $\tau_s(\infty)$ for different
values of $\epsilon$ could be obtained, for example we have
$\tau_s(\infty)_{\epsilon=0.1}=1.25\pm 0.06$ and
$\tau_s(\infty)_{\epsilon=0.4}=1.24\pm 0.06$. Again system shows
finite size scaling and is critical, but there is not a remarkable
variations in the exponents like $\tau_s$ as $\epsilon$ is changed.
Possibly they belong to identical universality class.

\subsection{One-site height probability}
In CASM with no asymmetries, one is able to find the probability of
having any height in the interval $[0,1]$. This is done using the
burning algorithm\cite{MajDharJPhysA}: consider the recurrent
configurations with the property that the height of the site $i$ is
in $[0,1]$. As this site burns after all of its neighbors, it should
be a leaf in the corresponding spanning tree. Therefore, the total
number of such configurations could be obtained by counting the
number of spanning trees in which site $i$ is a leaf. In the
modified model the same arguments holds for $h\in[0,1-\epsilon)$,
because the burning test still works and any site with
$h\in[0,1-\epsilon)$ has to be a leaf in the corresponding spanning
tree. Note that as we have links with weight $1+\epsilon$, the sites
with $h\in[1-\epsilon,1+\epsilon]$ may not be a leaf.

Similar to the way we find the probability of one in ASM, we modify
the the weights of the toppling matrix to ensure that the site $i$
is a leaf. This modification of weights is only done on a finite
number of lattice bonds, therefore the probabilities corresponding
to these local restrictions can be calculated in terms of finite
dimensional determinant.

\begin{figure}[b]
\begin{picture}(200,100)(0,0)
\includegraphics{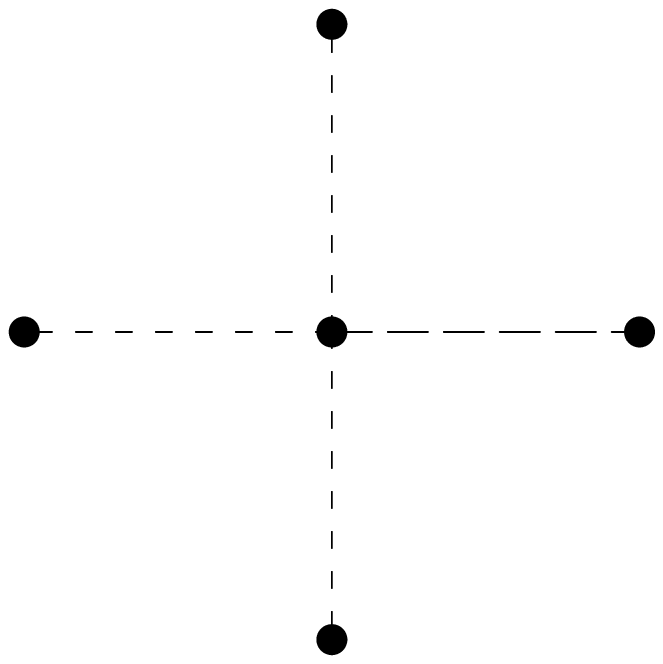}
\includegraphics{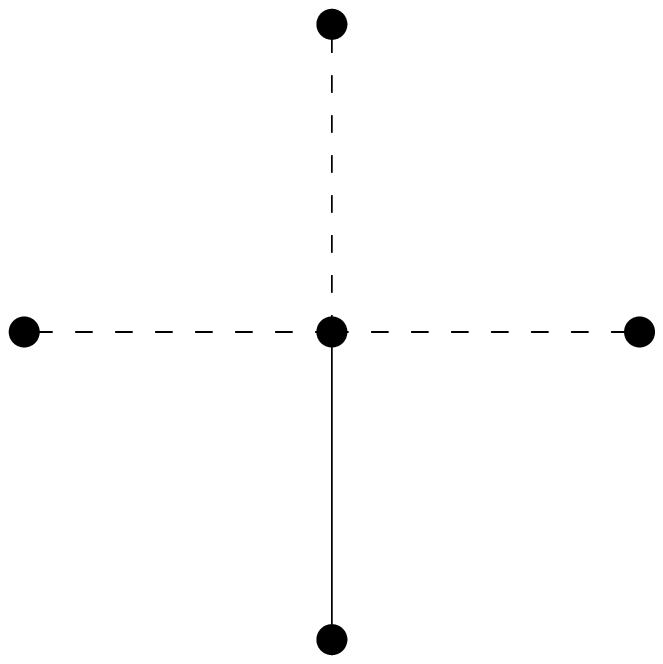}
\put(120,-4){A1} \put(324,-4){A2}
\end{picture}
\caption{Two types of bond modification}
\end{figure}

We modify the original model by removing the bonds to three
neighbors of site $i$ such that only one bond connects site $i$ to
the system. All four possible choices to cut the bonds are not
equivalent, there are two different modifications: $A_1$ and $A_2$
shown in Figure 6. In the case $A_1$, we delete all neighbor bonds
of site $i$ except a left or a right one which has weight
$1+\epsilon$, and change the weight of this remaining bond to
$1-\epsilon$ to ensure the height of site $i$ won't be more than
$1-\epsilon$. In the case $A_2$, one of the vertical bonds with the
weight $1-\epsilon$ remains unchanged and connects site $i$ to
system.

The two new toppling matrices are of the form
$\Delta^{(1)}=\Delta-B^{(1)}$ and $\Delta^{(2)}=\Delta-B^{(2)}$
respectively where  $B^{(1)}$ and $B^{(2)}$ are defect matrices:

\begin{equation}\label{matrixb2}
B^{(1)}=\left(%
\begin{array}{ccccc}
  3+\epsilon & -1+\epsilon & -1-\epsilon & -1+\epsilon & -2\epsilon \\
  -1+\epsilon & 1-\epsilon & 0 & 0 & 0\\
  -1-\epsilon & 0 & 1+\epsilon & 0 & 0 \\
  -1+\epsilon & 0 & 0 & 1-\epsilon & 0 \\
  -2\epsilon & 0 & 0 & 0 & 2\epsilon \\
\end{array}%
\right)
\end{equation}
\begin{equation}\label{matrixb}
 B^{(2)}=\left(%
\begin{array}{cccc}
  3+\epsilon & -1-\epsilon & -1+\epsilon & -1-\epsilon \\
  -1-\epsilon & 1+\epsilon & 0 & 0 \\
  -1+\epsilon & 0 & 1-\epsilon & 0 \\
  -1-\epsilon & 0 & 0 & 1+\epsilon \\
\end{array}%
\right),
\end{equation}
As we have left-right and up-down symmetries, the number
configurations where $h_i\in[0,1-\epsilon)$ is proportional to
$N=2(\det \Delta^{(1)}+\det \Delta^{(2)})$. So the probability of
finding at most $1-\epsilon$ amount sand at a site is given by:
\begin{equation}\label{prob}
P(1-\epsilon)=\frac{N}{4\det\Delta}=\frac{1}{2}\det(I-B^{(1)}G)+\det(I-B^{(2)}G),
\end{equation}
where $G=\Delta^{-1}$, is Green function matrix with the following
integral form:

\begin{equation}\label{green}
G_{ij}-G_{00}=\int_{-\pi}^{\pi}\frac{dp_{i}}{2\pi}\int_{-\pi}^{\pi}\frac{dp_{j}}{2\pi}\frac{\cos(
i\, p_{i})\cos(j\, p_{j})-1}{4-2(1+\epsilon)\cos
p_{i}-2(1-\epsilon)\cos p_{j} }.
\end{equation}\vspace{1mm}

The new toppling rules with elliptic asymmetry in CASM make the
values of Green function to depend on the orientation of two sites
an addition to their distances. Hence their calculation is a bit
more tricky. For example, in the symmetric case it is enough to
calculate $G_{ii}$, all other components of $G$ could be obtained
easily thereafter. We have computed the Green functions analytically
and a few of them which were necessary to derive $P(1-\epsilon)$ are
listed in Appendix. Using these Green functions, the determinants
can be computed and $P(1-\epsilon)$ is derived. The formula is very
long expression and we do not bring it here, we have sketched it in
Figure 7. The result is very good in agreement with the simulations
done. As we see, for $\epsilon=0$ we arrive at the well-known result
$P(1)$ in BTW model. The probability gradually comes down till it
vanishes at $\epsilon=1$ which could be expected.
\begin{figure}[b]
\begin{picture}(200,105)(0,12)
\includegraphics{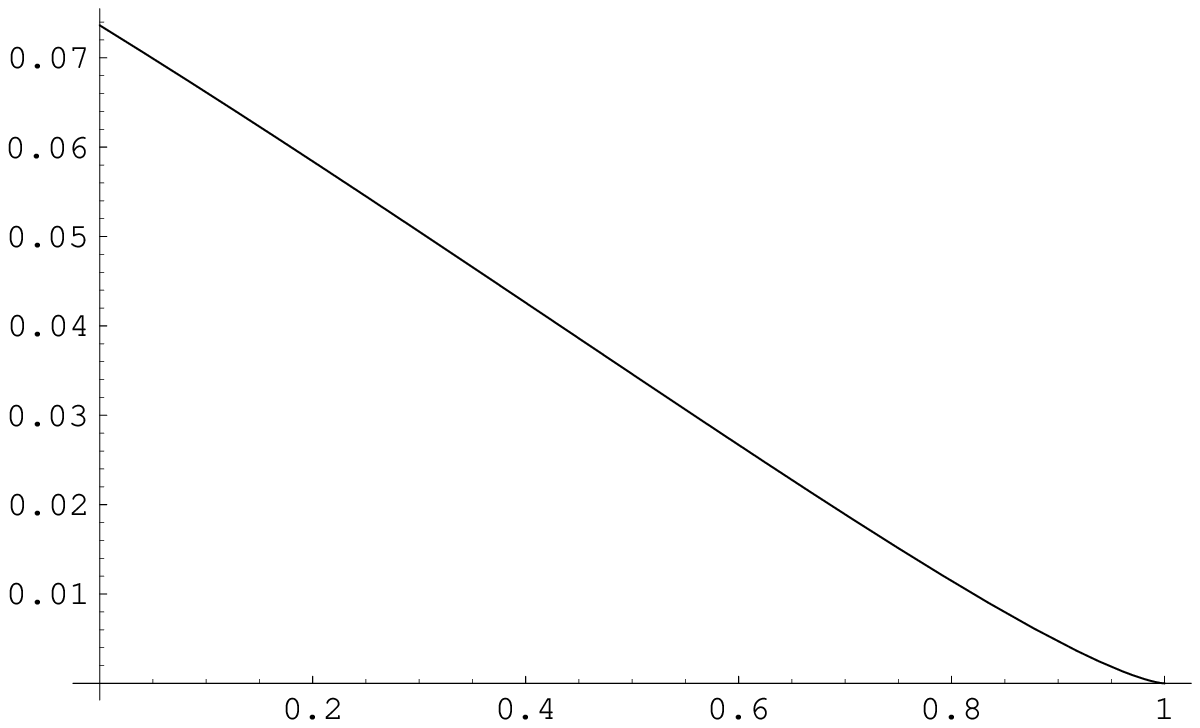}
\put(95,120){$P(1-\epsilon)$} \put(319,-4){$\epsilon$}
\end{picture}
\caption{The probability $P(1-\epsilon)$ as a function of
$\epsilon$}
\end{figure}
It is also possible to calculate other one site probabilities such
as the probability $h\in[1-\epsilon,1+\epsilon)$. In these cases,
you have some non-local constraints, just like higher-than-one one
site probabilities in usual ASM.
\subsection{Two-point correlation function}
Again, using the same scheme introduced by \cite{MajDharJPhysA}, one
is able to find correlation functions of finding sites with height
less that $1-\epsilon$. Making the modifications according to
configurations $A_1$ and $A_2$ in the vicinity of any two sites, it
is possible to find the probability that both sites have  heights
less than $1-\epsilon$. In order to compute this joint probability,
we consider one site at the origin and another at $i=(r,\theta)$.
There are four different cases depending on how we modify the
toppling rules: the defect matrices associated with either of the
site could be $B^{(1)}$ or $B^{(2)}$. Collecting all the four
possible configurations, one finds:

\begin{equation}\label{probtwo}
P(1-\epsilon,1-\epsilon)=\frac{1}{4}\left(\sum_{n,\acute{n}=1}^{2}
\det\left(I-\left(%
\begin{array}{cc}
   G_{00}& G_{0i} \\
  G_{i0} & G_{ii} \\
\end{array}%
\right)\left(%
\begin{array}{cc}
  B^{(n)} & 0 \\
  0 & B^{(\acute{n})} \\
\end{array}%
\right)\right)\right).
\end{equation}

The $G$ blocks denote green function matrixes at two sites and
$G_{0i}=(G_{i0})^{t}$. The short distance Green functions are given
in Appendix, therefore to compute the above determinants  at the
scaling limit we have to know the expansion of Green function at far
distances. This could be obtained by solving the deformed continuum
Laplace equation:

\begin{equation}\label{greenexpansion}
G(r,\theta)=\frac{1}{\sqrt{1-\epsilon^{2}}}\left(-\frac{1}{2\pi}\ln
r -\frac{1}{4\pi}\ln
\left(\frac{\cos^{2}\theta}{1+\epsilon}+\frac{\sin^{2}\theta}{1-\epsilon}\right)-\frac{\gamma}{2\pi}-\frac{\ln
8}{4\pi}+\frac{1}{24\pi
r^{2}\left(\frac{\cos^{2}\theta}{1+\epsilon}+\frac{\sin^{2}\theta}{1-\epsilon}\right)}+\ldots\right)
\end{equation}
where $\gamma=0.577\ldots$ is the Euler-Mascheroni constant.
Using all these together, we are ready to find the two point
probability:
\begin{equation}\label{two}
P(h_0<1-\epsilon,
h_r<1-\epsilon)=\left(P(1-\epsilon)\right)^2+\frac{f(\epsilon,\theta)}{r^{4}}+\ldots
\end{equation}
where $f$ is derived as an analytic functions in terms of $\epsilon$
and $\theta$. Again it is a very long expression and we do not bring
it here. We have sketched the function $f$ in polar coordinate for
two different values of $\epsilon=0.1$ and $0.2$ in Figure 8. The
result shows that the correlations are long ranged and have scaling
property, though there is no rotational symmetry, and the amount of
this anisotropy depends on $\epsilon$.

\begin{figure}[t]
\begin{picture}(200,100)(0,0)
\includegraphics{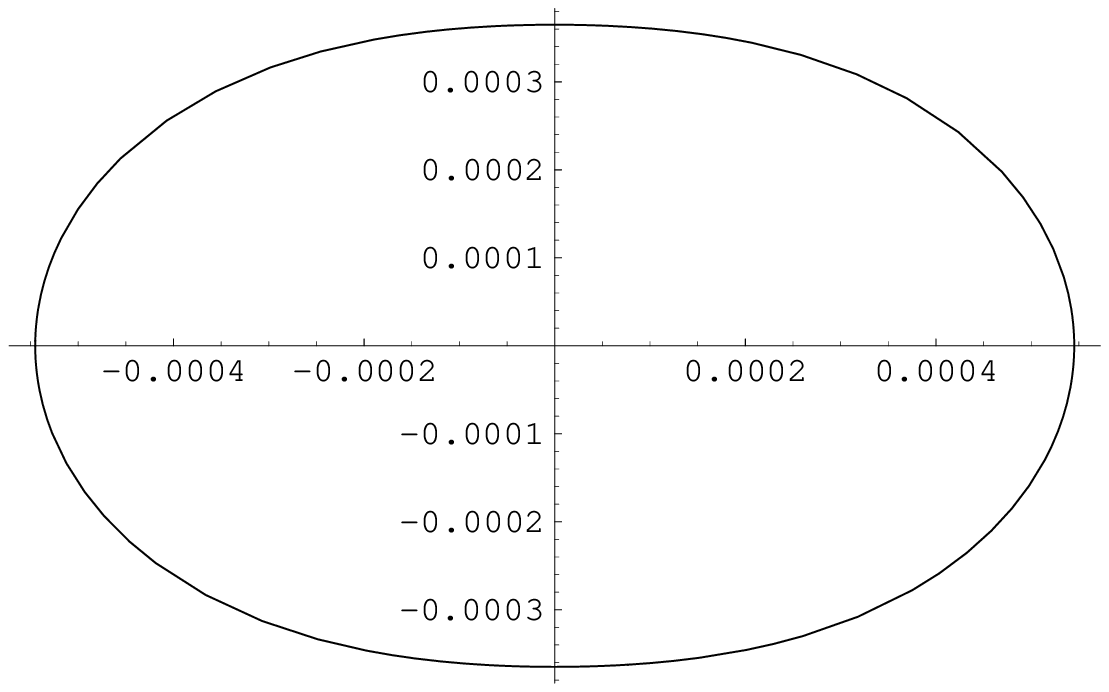}
\includegraphics{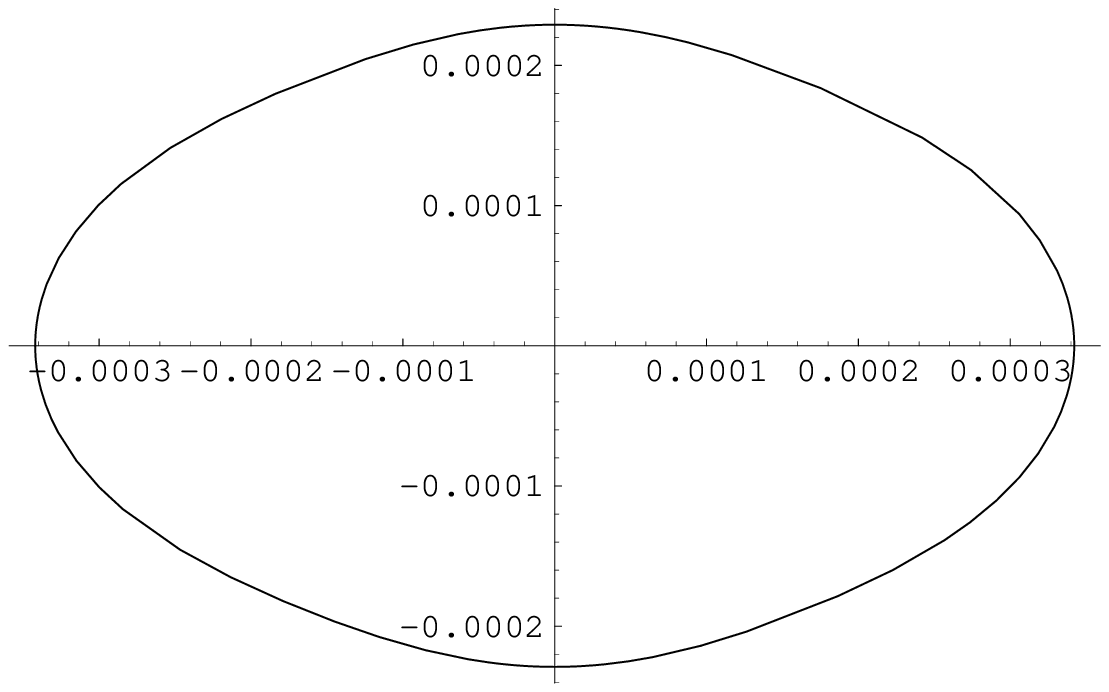}
\put(88,0){$\epsilon=0.1$}\put(312,0){$\epsilon=0.2$}
\end{picture}
\caption{The amount of rotational asymmetry of two point functions
for $\epsilon=0.1$ and $0.2$.}
\end{figure}

\subsection{Boundary effects on height $1-\epsilon$ probability}
As boundary sites play important role in sandpile models, it would
be interesting to consider the model in presence of different
boundary conditions (BCs). We consider two types of boundary
conditions, open and closed \cite{Brankov}, according to toppling
rules in our model. In the case of open BC, we set $\Delta_{ii}=4$
for boundary sites. This means that the amount of sand leaving the
system when a boundary site topples is $1-\epsilon$ or $1+\epsilon$
depending on whether the site is on a horizontal or vertical edge.
For closed BC we set $\Delta_{ii}=3+\epsilon$ on horizontal edge and
$\Delta_{ii}=3-\epsilon$ for vertical edge so that there would be no
dissipation on the boundary. It maybe hard to keep track of vertical
and horizontal edges, an easier way is to consider
$\epsilon\in(-1,1)$, so the transformation
$\epsilon\rightarrow-\epsilon$ would interchange vertical an
horizontal edges.

\begin{figure}[t]
\begin{picture}(100,130)(0,0)
\includegraphics{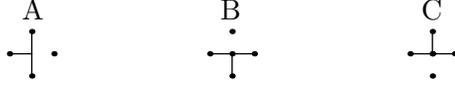} \put(143,114){A}\put(218,114){B}\put(294,114){C}
\end{picture}
\caption{Three different configurations to modify the toppling
matrix near a boundary}
\end{figure}

For simplicity, we consider that the model is defined on the upper
half plane and the open/closed boundary is located at $y=1$. Let's
compute the probability of having an amount of sand less than
$1-\epsilon$ at site $i=(0,y)$ in the upper half plane. To this end
we have to compute the open and closed Green functions for two sites
$i_{1}=(x_{1},y_{1})$ and $i_{2}=(x_{2},y_{2})$. This could be done
using the image method:
\begin{eqnarray}\label{openclosed}
G^{op}(x_{1},y_{1};x_{2},y_{2})&=&G(x_{1}-x_{2},y_{1}-y_{2})-G(x_{1}-x_{2},y_{1}+y_{2}) \\
G^{cl}(x_{1},y_{1};x_{2},y_{2})&=&G(x_{1}-x_{2},y_{1}-y_{2})+G(x_{1}-x_{2},y_{1}+y_{2}-1)
\end{eqnarray}
The calculations of height $1-\epsilon$ probability in the presence
of the boundary are similar to that on the plane. There are three
different configurations as shown in Figure 9, therefore\footnote{It
turns out that the configurations $B$ and $C$ have the same
contribution at large distances.}
\begin{equation}\label{probboundary}
P^{op,cl}(1-\epsilon,y)=\frac{1}{4}\left(2~\det\left(I-B^{(1)}G^{op,cl}_{h,A}\right)+\det\left(I-B^{(2)}G^{op,cl}_{h,B}\right)
+\det\left(I-B^{(2)}G^{op,cl}_{h,C}\right)\right).
\end{equation}
Using Eq. (\ref{greenexpansion}) we are able to compute the
determinants for distances far from boundary, then we find that the
probability for height $1-\epsilon$ at a distance $y$ from boundary
is:
\begin{equation}\label{probboundary2}
P^{op,cl}(1-\epsilon,y)=P(1-\epsilon)\pm\frac{S(\epsilon)}{y^{2}}+O\left(\frac{1}{y^{4}}\right),
\end{equation}
where  the constant term $P(1-\epsilon)$ is bulk probability for a
site to have height at most $1-\epsilon$ and is given by Eq.
(\ref{prob}). The leading term that depends on the distance from
boundary falls off as $1/y^{2}$ with coefficient $S(\epsilon)$ which
is a complicated analytic function of $\epsilon$ and is sketched in
Figure 10. Plus sign corresponds to open boundary condition and
minus sign to closed ones. It can be shown that this probability at
distance $x$ from a vertical open or closed boundary has a similar
behavior.

\begin{figure}[t]
\begin{picture}(200,100)(0,0)
\includegraphics{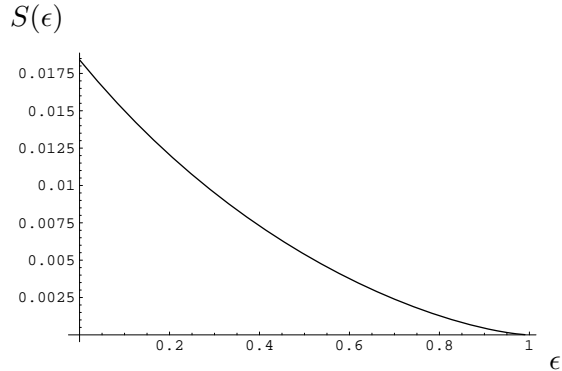}
\put(115,130){$S(\epsilon)$} \put(319,0){$\epsilon$}
\end{picture}
\caption{ $S(\epsilon)$ as a function of $\epsilon$}
\end{figure}

\subsection{Field Theory Description}

As mentioned before, the number of different recurrent
configurations of sandpile model (the partition function) is given
by determinant of the toppling matrix $\Delta$ which leads to the
action of $c=-2$ logarithmic conformal field theory (Eq.
(\ref{c-2action})). We rewrite the action in $x$-$y$ coordinates:
\begin{equation}\label{action}
S=\int
-\theta\partial_{x}^{2}\bar{\theta}-\theta\partial_{y}^{2}\bar{\theta}.
\end{equation}
The lattice rotational symmetry is broken when we apply elliptic
asymmetry in toppling rules. So it is expected the field theory
describing our model not to be conformal invariant. The perturbed
action turns out to be
\begin{equation}\label{actionasy}
S\sim\int
-\theta\partial_{x}^{2}\bar{\theta}-\theta\partial_{y}^{2}\bar{\theta}
-\epsilon(\theta\partial_{x}^{2}\bar{\theta}-\theta\partial_{y}^{2}\bar{\theta}),
\end{equation}
which is rotationally asymmetric.

In this field theory, following the grassmanian method used
in\cite{MRR}, it is possible to assign a scaling field to height
variable being in  $[0,1-\epsilon)$. The probability of height
$1-\epsilon$ can be written as
\begin{equation}\label{field}
P(1-\epsilon)=\frac{\int d\theta_{i}\int
d\bar{\theta}_{j}\exp\left(\sum\theta_{i}\Delta_{ij}\bar{\theta}_{j}\right)
\frac{1}{2}\left(\exp\left(-\theta_{i}B^{(1)}_{ij}\bar{\theta}_{j}\right)+\exp\left(-\theta_{i}B^{(2)}_{ij}\bar{\theta}_{j}\right)\right)}
{\int \d\theta_{i}\int
d\bar{\theta}_{j}\exp(\sum\theta_{i}\Delta_{ij}\bar{\theta}_{j})},
\end{equation}
which resembles the expectation value of the field
$(1/2)\left(\exp(-\theta B^{(1)}\bar{\theta})+\exp(-\theta
B^{(2)}\bar{\theta})\right)$. Following \cite{MRR}, we can assign a
field to this height variable in the following way:
\begin{equation}\label{field2}
\phi(1-\epsilon)=\frac{1}{2}\langle\langle
\exp(-\theta_{i}B^{(1)}_{ij}\bar{\theta}_{j})+\exp(-\theta_{i}B^{(2)}_{ij}\bar{\theta}_{j})
\rangle\rangle,
\end{equation}
where $\langle\langle \ldots \rangle\rangle$ means that we have to
contract all $\t$'s and $\tb$'s except two. In the contraction we
have to use the Green functions given in appendix. Doing all this
and going to continuum limit we will arrive at
\begin{equation}\label{fieldend}
\phi(1-\epsilon)=-\left(f(\epsilon)\partial_{x}\theta\partial_{x}\bar{\theta}+g(\epsilon)\partial_{y}\theta\partial_{y}\bar{\theta}\right)
\end{equation}
$f(\epsilon)$ and $g(\epsilon)$ are complicated functions with long
terms that are plotted in Figure 11.

\begin{figure}[t]
\begin{picture}(200,100)(0,0)
\includegraphics{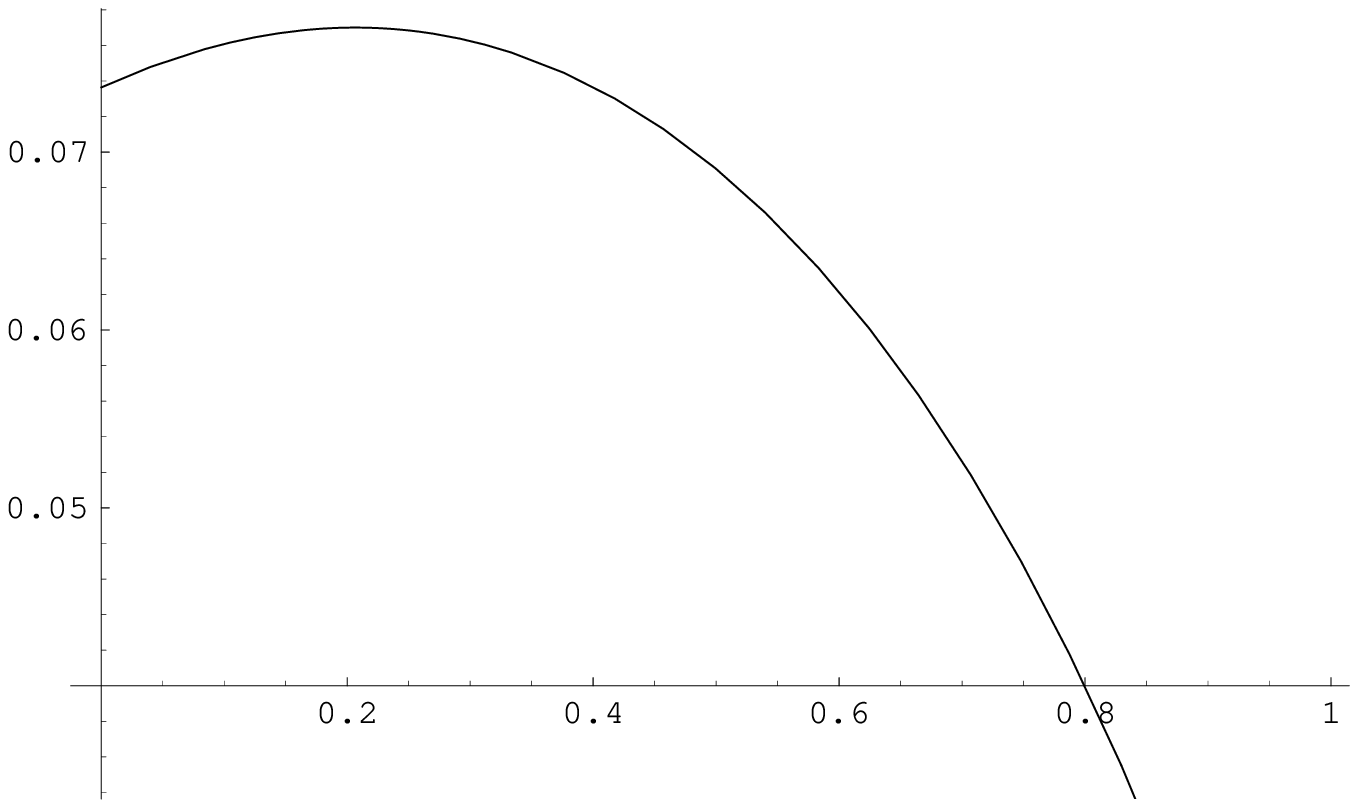}
\includegraphics{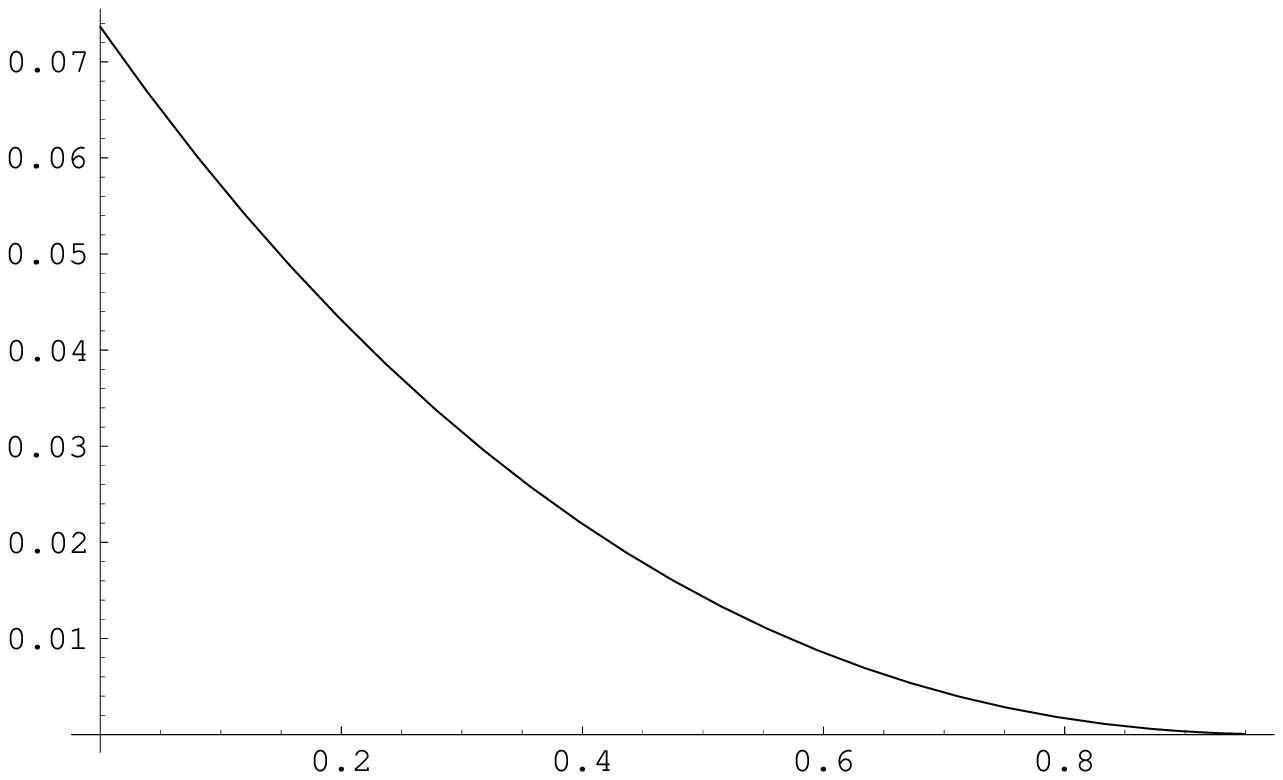}
\put(10,110){$f(\epsilon)$}
\put(224,110){$g(\epsilon)$}\put(210,5){$\epsilon$}\put(415,5){$\epsilon$}
\end{picture}
\caption{$f(\epsilon)$ and $g(\epsilon)$ are sketched as functions
of $\epsilon$.}
\end{figure}

As we see, setting $\epsilon=0$ we will arrive at the known field in
ASM. As we increase $\epsilon$ the function $g$ gradually approaches
to zero and is always positive. However the function $f$ increases
up to $\epsilon\simeq 0.2$ and then comes down becomes zero at
$\epsilon=1$. This is natural, because the limit
$\epsilon\rightarrow 1$ means you want to find the field associated
with the probability of height $h$ be in the interval
$[0,1-\epsilon]=\emptyset$, therefore the field should be zero.

At any case it is very interesting to investigate this new field
theory. In this theory, the scale symmetry is preserved and we
expect criticality. However the rotational symmetry is broken and we
do not have the full conformal invariance, though the system is
still solvable from field theory point of view. Possibly on can
investigate this model in the context of perturbed conformal field
theory \cite{ABZ,RajabpourRouhani}. It can be argued that the action
(\ref{actionasy}) could become symmetric if you rescale horizontal
and vertical directions with suitable (and different) scale
parameters\cite{DharPri}, therefore this new theory should be
identical with the previous one. This argument, if applied to the
action is completely correct, and if one construct the field theory
through defining the correlation functions the two theory will be
identical. But if you ask questions about the field assigned to
physical properties, such as the probability of height of a site
being less that $1-\epsilon$, the situation becomes more complicated
and it is not easy to find an easy way to relate the corresponding
fields in the two theory. For example the field associated with the
probability of height of a site being less that $1-\epsilon$ in
ordinary CASM is $(1-\epsilon)\phi_1$ where $\phi_1$ is the field of
height one in ASM. However the same field in asymmetric case is
given by equation (\ref{fieldend}) and could not be derived from
$\phi_1$ via a simple rescaling. On the other hand it has been seen
that the scaling exponent $\tau_s$ is not sensitive to the value of
$\epsilon$. Putting all these together, it seems that the elliptical
asymmetry do not change the universality class of the theory.

It is also possible to restore the rotational symmetry
statistically. If we assume the on-site asymmetries have quenched
randomness; that is, on some sites $\epsilon$ takes positive values
and on some other it takes negative values, then there is no
preferred direction statistically. The model should be formulated
with more care, however it is very interesting to see if adding such
quenched randomness will take the system to a new fix point or not.
Work in this direction is in progress. \vspace{5mm}

{\Large \bf Acknowledgment}

We would like to thank D. Dhar for his
helpful comments and careful reading of the manuscript.

\section{Appendix}
We collect some values of Green function $G$ that we have used
through the paper:
\begin{eqnarray*}\label{green}
G(1,1)&=&-\frac{1}{\pi\sqrt{1-\epsilon^{2}}}\\
G(0,1)&=&\frac{\arcsin\left(\sqrt{\frac{1-\epsilon}{2}}\right)}{\pi(\epsilon-1)}\\
G(1,0)&=&-\frac{(\pi-2{\rm
arccotan}(\frac{1+\epsilon}{\sqrt{1-\epsilon^{2}}}))}{2\pi(1+\epsilon)}\\
G(0,2)&=&-\frac{2(1+\epsilon)}{\pi(-1+\epsilon)\sqrt{1-\epsilon^{2}}}-\frac{4\arcsin\left(\sqrt{\frac{1-\epsilon}{2}}\right)}{\pi(-1+\epsilon)^{2}}\\
G(2,0)&=&\frac{2\left(\sqrt{1-\epsilon^{2}}-\pi+2{\rm
arccotan}\left(\sqrt{\frac{1+\epsilon}{1-\epsilon}}\right)\right)}{\pi\left(1+\epsilon\right)^{2}}
\end{eqnarray*}

\end{document}